\documentclass[aps,prl,twocolumn]{revtex4}

\usepackage{graphicx}
\usepackage{amssymb}
\usepackage{enumerate}

\input epsf

\begin{document}

\title{Photon Counting for Bragg Spectroscopy of Quantum Gases}
\author{J. M. Pino}
\author{R. J. Wild}
\author{P. Makotyn}
\author{D. S. Jin}
\author{E. A. Cornell}

\address{JILA, Quantum Physics Division, National Institute of Standards and Technology
and Department of Physics, University of Colorado, Boulder, Colorado
80309-0440, USA}
\date{\today}

\begin{abstract}
We demonstrate a photon-counting technique for detecting Bragg excitation of an ultracold gas of atoms. By measuring the response of the light field to the atoms, we derive a signal independent of traditional time-of-flight atom-imaging techniques. With heterodyne detection we achieve photon shot-noise limited detection of the amplification or depletion of one of the Bragg laser beams. Photon counting for Bragg spectroscopy will be useful for strongly interacting gases where atom-imaging detection fails. In addition, this technique provides the ability to resolve the evolution of excitations as a function of pulse duration.

\end{abstract}



\maketitle\

\section{Introduction}

Ultracold quantum gases provide clean, controllable model systems for exploring quantum many-body physics \cite{Bloch2008}, and a powerful probe of these interacting quantum systems is the spectroscopy of low-energy excitations. For example, intriguing behaviors such as superfluidity, fermion pairing, and exotic weakly bound molecules can be revealed in their characteristic excitation
spectra. In many cases, measured spectra can be directly compared with many-body theory to test our understanding of these rich systems.

Any experiment designed to probe excitations requires the same
essential components: a way to excite the sample and a way to measure the response. To drive the excitations one applies a field to perturb the gas, and spectra are obtained by measuring the response of the ultracold gas as a function of the driving-field frequency. In ultracold atom experiments, this response is measured by imaging the atom cloud after the perturbation, and observing the response as some change in the density distribution of the imaged cloud.

There exists, however, in any excitation spectroscopy, an  alternative way to measure the response of the quantum gas to a driving field. Just as the quantum gas has responded to the probe field, the probe field must have an equivalent response to the quantum gas. In this paper, we present  a technique to measure this complementary information in Bragg spectroscopy by measuring the change in the number of photons in one of the laser fields used to drive the Bragg excitations. While in this work we use Bragg spectroscopy,  this technique may more generally be applicable to other excitation spectroscopies as well \cite{Peden2009a}. For ultracold atoms where the number of atoms is typically small ($\sim 10^5$), the challenge, of course, is to have adequate signal-to-noise in measuring photon number to detect the Bragg response. For smallish atom samples, this will typically require a photon shot-noise limited measurement. To that end, we detail the experimental setup and the techniques of a heterodyne-based detection scheme used to realize shot-noise-limited photon counting. We also explore the advantages and limitations of this technique.

To motivate tackling the challenge of photon counting, we point out that this technique avoids some of the pitfalls inherent to spectroscopy measurements where the cloud is imaged. For example, strong interactions, which complicate the response seen in the density distribution of a Bose-Einstein condensate (BEC) \cite{Papp2008a}, provide no additional complications for the photon counting approach.  Indeed, a primary motivation for this work is our desire to extend Bragg spectroscopy of BEC to the case where we have both strong interactions and low momentum excitations. Finally, we present a powerful new feature of our technique, which is that we can directly probe the time evolution of the excitation process, even during the course of a single laser pulse.

\section{Bragg Spectroscopy}

Conceptually, a Bragg excitation can be thought of as a coherent scattering process involving absorption of a photon from one of the Bragg beams and stimulated emission into the other: a two-photon transition. It is this process of either absorption or emission that we aim to measure with our photon-counting technique. The process leaves the excited atoms in the same internal state, but with a new momentum, $\textbf{k}$, determined by the geometry and wavelength of the Bragg beams. The two lasers have slightly different frequencies, to account for the energy of the excitation, and we vary this frequency difference, $\omega$, for our spectroscopy.

Typically, the Bragg response is measured by looking at the cloud using time-of-flight atom imaging, where the gas is suddenly released from the trap and allowed to expand  before imaging. Bragg excitations in a weakly interacting BEC then appear as atoms in a distinct cloud outside of the main cloud. The position of the new cloud, which is seen after the expansion from the trap, reflects the momentum of the excitation.  This new cloud's density reflects the strength of the excitation. Pioneering studies of weakly interacting BECs have been done with such measurements \cite{Kozuma1999a,Stenger1999b,Steinhauer2002a}, however, there are limitations. In particular, in the case of strong interparticle interactions \cite{Papp2008a,Veeravalli2008} or low-momentum excitations \cite{Steinhauer2002a}, the excitations no longer appear as a clearly distinguishable second cloud in the momentum-space image and the response becomes more difficult to quantify.

\begin{figure}[h]
\begin{center}
\includegraphics[width=80mm]{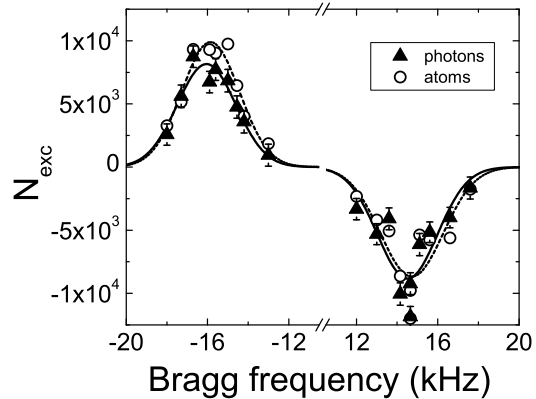}
\caption{Bragg spectrum of a $^{85}$Rb BEC of $5.5\times10^4$ atoms at k = 16 $\mu$m$^{-1}$, measured in two different ways. The horizontal axis shows the frequency difference between the Bragg beams. The vertical scale shows the number of excitations, which is measured using traditional absorption imaging of ejected atoms (hollow circles), as well as with the photon-counting technique presented here (black triangles). The error bars on black points represent the shot noise in the photon counting measurements. The photon counting measurements used three pulses of equal length. The first and the third pulses used only a single weak beam to make an average background measurement, with no Bragg excitation. During the second pulse, both Bragg beams illuminate the condensate to induce Bragg scattering and we subtract the averaged background to count the number of photons gained or lost in the weak beam due to the Bragg excitation.
}
\label{fig:lineshape}
\end{center}
\end{figure}

We show a measured Bragg lineshape for a weakly interacting $^{85}$Rb BEC in Fig.\ \ref{fig:lineshape}. On the horizontal axis, we have the frequency difference between the two Bragg beams, which sets the energy of the Bragg excitation. On the vertical axis, we have the Bragg signal, namely, the number of excitations due to the Bragg process. We define this signal such that it can be either positive or negative, reflecting the direction of the momentum transfer. The number of excitations due to the Bragg pulse are counted in two different ways, and one can see that the photon counting and the time-of-flight imaging signals agree well with each other. The two sets of data were acquired simultaneously, with each cycle of the experiment providing both a photon-counting and an atom-imaging data point. This demonstrates the complementary nature of the two techniques. The lines in Fig.\ \ref{fig:lineshape} are individual fits of the Bragg spectrum to two Gaussian functions. These fits can be used to extract a center frequency and an RMS width. In the rest of the paper we describe in detail the photon-counting technique demonstrated here.

A two-photon Rabi frequency characterizes the Bragg response and gives us an understanding of how the excitation rate will scale with the various experimental parameters. The two-photon Rabi frequency is given by
\begin{equation}
\Omega_{\scriptsize \textrm{Bragg}} = \frac{\Omega_{\scriptsize \textrm{weak}} \Omega_{\scriptsize \textrm{strong}}}{2\Delta_{\scriptsize \textrm{int}}}
\end{equation}
where $\Omega_{\scriptsize \textrm{weak}}$ and $\Omega_{\scriptsize \textrm{strong}}$ refer to the individual Rabi frequencies associated with each laser field, labeled here as weak and strong (the Bragg beam to be measured is the weak beam, for reasons that will be clear later). The detuning, $\Delta_{\scriptsize \textrm{int}}$, in rad/s, refers to the detuning of the beams from the atomic transition, and is much larger than $\omega$, the small frequency difference between the two beams. Using the Rabi frequency, we can define an excitation rate, $\Gamma$($\omega$) where
\begin{equation}\label{eq:BraggScattering}
\int \Gamma(\omega) d\omega = \frac{\pi}{2}S(\textbf{k})\Omega_{\scriptsize \textrm{Bragg}}^{2}
\end{equation}
here $S(\textbf{k})$ is the static structure factor, and gives the strength of the Bragg resonance for a particular $\textbf{k}$. \cite{Brunello2001a, Ketterle2001a}.

In this paper, we describe photon counting detection for Bragg spectroscopy of BECs of Rb atoms. In our experiment, we can create nearly pure BECs of either $^{85}$Rb or $^{87}$Rb atoms, with a final number of roughly $4\times10^4$ or $4\times10^5$, respectively, in a roughly 10 Hz spherical magnetic trap. Thanks to a magnetic-field Feshbach resonance at 155 G, we can study strongly interacting condensates using $^{85}$Rb atoms. The $^{87}$Rb condensates, on the other hand, have far greater atom number and give us a robust tool for characterizing the photon-counting technique with an increased signal-to-noise ratio.

Regardless of the method used to measure a Bragg response, adequate signal-to-noise will always be a prerequisite. The signal is set by the number of Bragg excitations, which is typically less than 10$\%$ of the total atoms in the sample in order to probe the linear response. For our $^{85}$Rb condensates, that would be some 4,000 excitations. If we allow for $10^5$ photons in the weak beam, this corresponds to a signal-to-noise ratio (SNR) of only 9 on resonance. This assumes a detector with perfect quantum efficiency as well as a shot-noise limited detection scheme.

For the best signal-to-noise ratio, assuming that a shot-noise limited measurement is available, we look to minimize the total number of photons in the weak beam. Keeping the excitation rate (and hence the signal) constant when decreasing the intensity of the weak beam, necessitates increasing the intensity of the strong beam (see Eq.\ \ref{eq:BraggScattering}). However, single-photon processes, namely off-resonant scattering of the stronger beam, limit the maximum permissible intensity in the strong beam, and this, in turn, limits the minimum intensity in the weak beam for a desired excitation rate. Note that having a large mismatch between the two Bragg beam intensities is a distinct requirement for photon counting compared to the usual atom-response detection, where there is no reason not to have equal intensities in the two beams.

Another limit to the minimum photon number in the weak beam is set by the spatial profile of the weak beam at the atoms. We tailor this spatial profile by focussing the weak beam onto the atoms, attempting to match the transverse spatial profile of the weak beam to that of the condensate in order to minimize the number of photons that would never interact with the condensate, and only add to the shot-noise of the measurement. In minimizing the number of weak beam photons, one could easily enter a regime where the number of Bragg photons scattered is a significant fraction of the weak beam photons themselves. In this regime, the transverse spatial profile of the weak beam intensity would be modified due to the scattering of Bragg photons out of (or into) the weak beam; this could, in effect, burn a ``hole" in the probe. To simplify the interpretation of the Bragg response, we keep the total number of photons in the region of overlap between the weak beam and the BEC large compared to the number of excitations. This provides another limit on  the minimum number of weak beam photons and, therefore, the maximum signal-to-noise ratio for the measurement.

In order to then make shot-noise limited measurements of the weak light, one could consider using detectors with integrated cascading amplification, such as avalanche photodiodes and photo multiplier tubes. However, in considering this option for our purposes, we found that avalanche photodetectors suffer from pulse-pileup for the powers we use, while photo multiplier tubes suffer from low quantum efficiencies in the IR, which at $5\%$ or so are too low to be useful. Therefore we consider the alternative of using detectors in their linear mode of detection. Silicon photodetectors have quantum efficiencies as high as $90\%$ and capacitances on the order of 3 pF. However, for the pulse lengths we seek ($\sim 1$ ms) the Johnson noise associated with the necessary feedback resistor corresponds to the shot noise of a 1 $\mu$W beam at 780 nm and therefore makes direct shot-noise-limited measurements impossible  for the powers used in these experiments. Our solution to these challenges is one as old as the AM radio: heterodyning.

\section{Heterodyne detection}

To motivate heterodyne detection, we consider the following scheme. We direct the weak Bragg beam onto a photodiode, and after some stages of amplification, directly measure a voltage, $S_{\scriptsize \textrm{direct}} $, proportional to the number of photons per second incident on the detector. To illustrate the role of shot-noise, we consider our signal to be proportional to the number of photons, $N_{\scriptsize \textrm{weak}}$, that hit the detector in some time $\tau$. In terms of an average voltage, the signal is $S_{\scriptsize \textrm{direct}} = e R \eta N_{\scriptsize \textrm{weak}} / \tau$, where $\eta$ is the quantum efficiency of the detector, $e$ is the charge of an electron, and $R$ is the transimpedance gain of the amplifier.

The signal-to-noise ratio, $S\!N\!R_{\scriptsize \textrm{direct}}$, on this direct weak beam detection is given by
\begin{equation}
S\!N\!R_{\scriptsize \textrm{direct}} = \frac{e R \eta N_{\scriptsize \textrm{weak}}/ \tau}{\sqrt{\delta_{\scriptsize \textrm{S}}^{2}+\delta_{\scriptsize \textrm{e}}^{2}}}
\end{equation}
where the shot-noise associated with the photocurrent is $\delta_{S} = (R e /\tau) \sqrt{\eta N_{\scriptsize \textrm{weak}}}$. Other sources of noise (dark noise from the detector, Johnson noise introduced in the amplification stages, noise on background light incident on the detector, etc.)\ will be referred to as electronic noise, and are represented by $\delta_{e}$. Shot-noise limited detection is defined as the regime where $\delta_{e}$ is an insignificant contribution to the total noise ($\delta_{e} \ll \delta_{S}$). In this case,
\begin{equation}\label{eq:perfect}
S\!N\!R_{\scriptsize \textrm{shot-noise}} = \sqrt{\eta N_{\scriptsize \textrm{weak}}}.
\end{equation}
This is difficult to achieve in direct detection, where for typical experimental values of $\tau = 1$ ms, $N_{\scriptsize \textrm{weak}} = 10^5$, $\eta = 85\ \%$, we might expect $\delta_{e} = 1000\times\delta_{S}$ for a silicon photodetectors with a high bandwidth, low-noise transimpedance preamplifier.

We use a heterodyne scheme to overcome these difficulties. The idea of heterodyne detection is to amplify the signal \textit{optically} before detection, so that electronic noise is of no consequence. We do this by measuring the beat of the weak beam against another, more intense beam, which we will refer to as the local oscillator ($LO$). Because the beat signal, $S_{\scriptsize \textrm{beat}}$, goes as the square root of the product of the intensities of the weak and the LO beam, we are left with a signal that is much stronger than that of the original, weak beam signal.

The advantage heterodyne offers is that one may arbitrarily increase the intensity of the LO beam, so that the shot-noise from the LO photons dominates the electronic noise, and the total noise is given by
\begin{equation}\label{eq:LO_shot}
\sqrt{\delta_{S}^{2}+\delta_{e}^{2}} \simeq \delta_{S} = (e R  /\tau) \sqrt{\eta N_{\scriptsize \textrm{LO}}}
\end{equation}
where $N_{\scriptsize \textrm{LO}}$ is the number of photons from the LO beam during a time $\tau$.
Because increasing the LO beam intensity also increases the optical gain, we are left with a SNR on $N_{\scriptsize \textrm{weak}}$ using heterodyne detection as
\begin{equation}\label{eq:SNRweak}
S\!N\!R_{\scriptsize \textrm{hetero}} = \mathcal{C} \sqrt{\eta N_{\scriptsize \textrm{weak}}}
\end{equation}
where the contrast, $\mathcal{C}$, is a number between zero and one that describes the quality of the mode-matching between the two heterodyne beams. One notes that for perfect contrast ($\mathcal{C} = 1$), $S\!N\!R_{\scriptsize \textrm{hetero}}$ is our stated goal for a shot-noise limited measurement as in Eq.\ \ref{eq:perfect}. These principles of optical heterodyning are well established, and we refer the reader to Refs.\ \cite{Saleh1991,Yariv1997,Andonovic1989} for a more thorough discussion.

Measurements of the LO shot-noise, $\delta_s$, also serves to calibrate the overall gain of our system, $R$. Provided one knows $\eta$ (which is readily available from the photodiode's datasheet) and has a calibration for $N_{\scriptsize \textrm{LO}}$ (which is straightforward for the relatively high power of the LO beam), one can rewrite Eq.\ \ref{eq:LO_shot} as

\begin{equation}
R = \delta_{S} \frac{(\tau / e)}{\sqrt{\eta N_{\scriptsize \textrm{LO}}}}.
\end{equation}

\section{Optical Layout and Electronics}

\begin{figure*}[t!]
\begin{center}
    \includegraphics[width=6in]{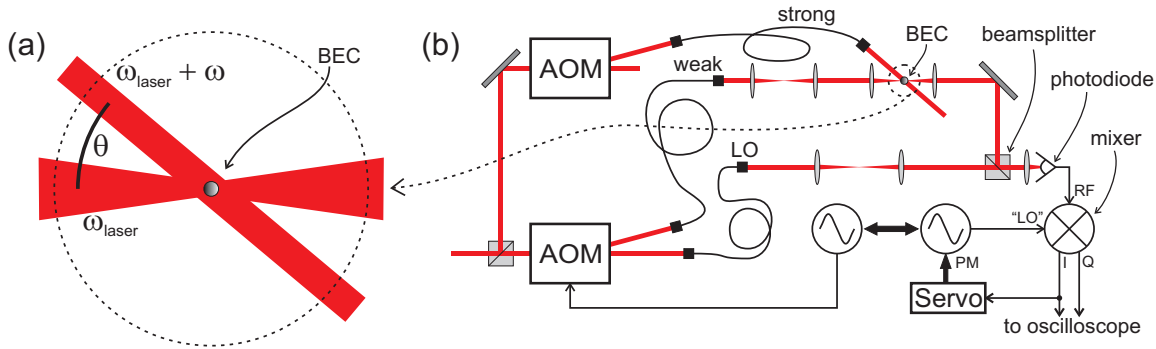}
        \caption{Diagram of the Bragg beams (a) and the heterodyne setup for Bragg scattering via photon counting (b). The Bragg beam frequencies ($\omega_{\scriptsize \textrm{laser}}, \omega_{\scriptsize \textrm{laser}}+\omega$) are offset by $\omega$, and create excitations at momentum $\hbar k = 2 \hbar k_{\scriptsize \textrm{laser}} \sin(\theta/2)$, where $\theta$ is the angle between the Bragg beams and $k_{\scriptsize \textrm{laser}} = 2\pi/780$ nm  in our case. In (b) the three beams (labeled weak, LO and strong) are derived from the same laser source and individually fiber coupled. The optics for the LO beam are chosen to give the same spatial mode as that of the weak beam. After being combined on a beamsplitter, the weak and LO beams illuminate a photodiode and the beat signal is sent to a demodulating mixer. The two quadrature outputs of the demodulator, I and Q, are sent to an oscilloscope for the measurement. Also illustrated in the figure is a servo loop that functions to minimize the phase fluctuations. The servo minimizes the demodulator output, I, by feeding back to a phase modulation input on the synthesizer that drives the ``LO" port of the mixer.  This synthesizer is phase-locked to the synthesizer driving the weak beam acousto-optic modulator (AOM). Not shown in the schematic is the synthesizer driving the strong-beam AOM, which is also phase referenced to the synthesizer driving the weak-beam AOM.
    }
\label{fig:layout}
\end{center}
\end{figure*}

While the principle behind shot-noise limited heterodyne detection is straightforward, the implementation of such a technique has a number of subtleties, which we lay out here. We illustrate both the optical and the RF design in Fig.\ \ref{fig:layout}. We begin our discussion on the optical side of things, and then consider the RF.

In Fig.\ \ref{fig:layout}(a), we show the two Bragg beams intersecting at the location of the BEC. The third, much more intense LO beam, shown in Fig.\ \ref{fig:layout}(b), functions only to amplify the weak beam for our heterodyne detection, and avoids the atoms altogether. All three beams are derived from the same laser, and are offset in frequency after passing through acousto-optical modulators (AOM). Each AOM has a dedicated synthesizer, with an additional synthesizer driving the mixer as well. All three of these synthesizers are phase-locked to each other to ensure coherence between the Bragg beams.

There are two challenges in optimizing this heterodyne interferometer for shot-noise limited detection. The first is achieving a contrast, $\mathcal{C}$, as close to unity as possible. In Eq.\ \ref{eq:SNRweak}, we see that the signal-to-noise ratio scales linearly with the contrast. Therefore, we would like the best possible mode-matching between the two beams, both spatially as well as in terms of polarization. In light of these two requirements, we launch the two beams into identical, polarization-maintaining fibers. The weak and the LO beams subsequently pass through different sets of lenses, which serve to match the focussed size of the weak beam to the size of the BEC, as well as to match the spatial mode of the LO beam to that of the weak when they recombine on the beamsplitter. To avoid interactions between the intense LO light and the atoms, the LO beam has its own optical path that avoids the vacuum chamber and the atoms. In addition, we have an optical isolator (not shown in figure) in between the beamsplitter and the photodiode to extinguish LO light backreflections off of the photodiode that interact with the atoms.

The second challenge is to minimize any jitter in the relative phase ($\phi$) between the two beams. As we will see later, this phase jitter introduces noise that limits us from achieving shot-noise limited detection for longer Bragg pulses. We find that the biggest source of instability in the relative phase is due to the optical fibers. These fibers are sensitive to mechanical vibrations and acoustical noise. We thus use relatively short fibers  (1 m) to minimize phase jitter between the two beams, while still providing the necessary filtering of the spatial mode.

Combining the two beams with a beamsplitter provides good spatial overlap, at the cost of weak beam attenuation from light lost at the unused beamsplitter port. This can be included in Eq.\ \ref{eq:SNRweak}, by  simply replacing $\eta$ with a lower effective quantum efficiency for the detection system. For the best signal-to-noise ratio, we would like to measure as much of the weak beam as possible. This must be balanced, however, with our need for large enough LO power to overwhelm the electronic noise. Typically, we work with 90\% reflection of the weak beam and 10\% transmission of the LO beam, corresponding to 250 $\mu$W of LO power on the photodiode.

To measure and maximize the contrast, we replace the light launched into the weak beam fiber with light at the same frequency and intensity as the LO. With this homodyne interferometer, one can easily assess the quality of the contrast by eye, by looking at the pattern of the interference fringes at the photodiode. Our procedure is to first align the beam such that the fringes, seen using a temporary lens to expand the beam, are  a circular pattern of rings. We then optimize the collimation of the LO beam by minimizing the total number of rings seen. We found this to be an essential step in getting good contrast. Once the beams are closely mode matched by eye, we perform the final steps of alignment by directly monitoring the DC output of the photodetector. We found it useful to modulate the phase of one of the beams, enough to wrap around 2$\pi$, so that the DC output swings between fully constructive ($\textmd{DC}_{\scriptsize \textrm{max}}$) and fully destructive ($\textmd{DC}_{\scriptsize \textrm{min}}$) interference. We are able to quantify the contrast through
\begin{equation}
\mathcal{C} = \frac{\textmd{DC}_{\scriptsize \textrm{max}}-\textmd{DC}_{\scriptsize \textrm{min}}}{\textmd{DC}_{\scriptsize \textrm{max}}+\textmd{DC}_{\scriptsize \textrm{min}}},
\end{equation}
true when the intensities of the two beams are the same.

We also take steps to minimize the electronic noise ($\delta_e$) that the LO's shot-noise must overcome. The relevant noise for the heterodyne measurement is that at the frequency of the beat note, $\Delta_{\scriptsize \textrm{LO}}$. Our photodiode circuit is designed to minimize the effects of inherent voltage noise at the op-amp input at this frequency ($2\pi \times 70$ MHz)  by way of a standard ``tank" circuit. The tank circuit consists of an inductor between the op-amp input and ground. The inductance is chosen so that when the photodiodes's internal capacitance is also considered the two make a resonant LC circuit at $\Delta_{\scriptsize \textrm{LO}}$. We are able to reduce the receiver (photodiode  plus amplifier \cite{SA5211}) dark noise to 2 pA/$\sqrt{\textrm{Hz}}$ which is typical for most commercially available transimpedance amplifiers. The subsequent stages of amplification are chosen so that the noise they add is small compared to that introduced in this first stage of amplification.


The voltage from the photodetector then goes to an RF mixer. As shown in Fig.\ \ref{fig:layout}(b), the mixer is a four-terminal device, called an I,Q-Demodulator; this is essentially two mixers in one, with the RF input split between the two. Port I is the same as the IF output of a standard mixer, while the Q port is the output of a second mixer whose ``LO" drive has a phase offset 90 degrees with respect to the phase of the I port's ``LO" drive. In our application, the outputs of the demodulator have the form
\begin{equation}\label{eq:QsumI}
S_{I} =(2 \mathcal{C} e R  \eta/\tau)\sqrt{N_{\scriptsize \textrm{LO}}N_{\scriptsize \textrm{weak}}}\cos{\phi}
\end{equation}
\begin{equation}\label{eq:QsumQ}
S_{Q} =(2 \mathcal{C} e R  \eta/\tau)\sqrt{N_{\scriptsize \textrm{LO}}N_{\scriptsize \textrm{weak}}}\sin{\phi}
\end{equation}
By summing the squares of $S_{I}$ and $S_{Q}$, we are able to measure the amplitude of the beat signal, regardless of the relative phase $\phi$. For a known LO beam power, $\propto N_{\scriptsize \textrm{LO}}/\tau$, our signal is directly proportional to the rate of weak beam photons, $\dot{n} = N_{\scriptsize\textrm{weak}}/\tau$.  We measure $S_{I}$ and $S_{Q}$ with a digitizing oscilloscope and perform the sum in subsequent software analysis.

While, in principle, we now have a signal that is insensitive to the relative phase $\phi$ between the weak beam and the LO beam, the demodulator is imperfect, due to non-linearities and voltage offsets in the I and Q outputs as well as deviations from perfect 90 degree phase offset. We therefore servo the phase of the RF driving the ``LO" port of the demodulator, as illustrated in Fig.\ \ref{fig:layout}(b). The servo minimizes $S_{I}$  by actively feeding back to the phase-modulation input of the relevant synthesizer, and this reduces the sensitivity of the quadrature sum to demodulation imperfections.

\section{Noise performance and Bragg dynamics}

\begin{figure}[h]
\begin{center}
\includegraphics[width=80mm]{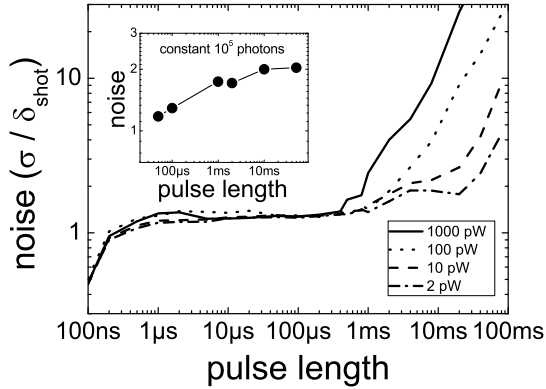}
    \caption{Noise performance of heterodyne detection for a LO power of 250 $\mu$W. Vertical scale is normalized to the shot-noise expected for the relevant pulse length and laser power. The legend shows the different weak beam powers used. Low-frequency drifts make our heterodyne scheme no longer shot-noise limited at long timescales. Inset has the same normalized vertical units, with the data shown at different weak beam intensities, corresponding to a constant $10^5$ photons.
    }
\label{fig:performance}
\end{center}
\end{figure}

In Fig.\ \ref{fig:performance}, we present data illustrating the shot-noise limited performance of our heterodyne detection. We measure noise by taking the standard deviation $\sigma$ in the measured weak beam photon number for $M$ consecutive pulses of equal length,
\begin{equation}
\sigma = \frac{1}{\sqrt{2}}\sum_{i=1}^{M-1}\sqrt{\frac{\left(N_{\scriptsize \textrm{weak}_{i+1}} -N_{\scriptsize \textrm{weak}_i}\right)^2}{M-1}}
\end{equation}
where $N_{\scriptsize \textrm{weak}_i}$ is the measured number of photons in the weak beam for pulse $i$. We plot the noise on our measurements of weak beam photon number, normalized to the expected shot-noise ($\delta_{\scriptsize \textrm{shot}} = \sqrt{\eta N_{\scriptsize \textrm{weak}}}  $) for that particular pulse duration and laser power, as a function of the duration of a fixed intensity pulse. In the inset of the figure, we plot the same noise measurement for the case where the weak beam intensity is varied to keep the total number of photons fixed at a constant $10^5$.

For a large range of pulse durations, the measurement is at, or within a factor of two of, the shot-noise limit. For pulses shorter than 1 $\mu$s the noise is artificially low due to an inline, low-pass filter. The increased noise at longer timescales sets an upper limit to the time available for our Bragg measurements and is probably caused by residual phase drift in our system. Servoing the laser power provided no significant improvement in the noise performance of our heterodyne detection.

\begin{figure}[h]
\begin{center}
\includegraphics[width=80mm]{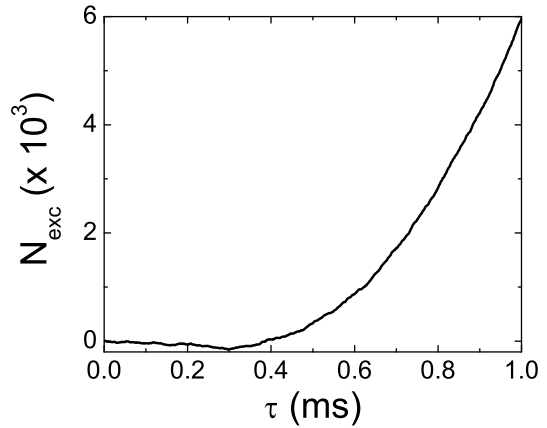}
    \caption{Bragg excitations as a function of time, measured with photon counting. The number of Bragg excitations is given by $N_{\scriptsize \textrm{exc}} = \int_0^{\tau} (\dot{n}(t)-\dot{n}_{\scriptsize \textrm{avg}}) dt$, where $\dot{n}_{\scriptsize \textrm{avg}}$ is the average rate of weak beam photons measured when no strong light is present. The Bragg pulse begins at 0 ms. These measurements were performed on resonance for a BEC of 400,000 $^{87}$Rb atoms.
    }
\label{fig:Bragg_v_time}
\end{center}
\end{figure}

A potentially useful feature of a photon-counting measurement is the ability to measure the dynamics of Bragg excitations during a single laser pulse. We demonstrate this capability in Fig.\ \ref{fig:Bragg_v_time}, where we  plot the number of excitations, $N_{\scriptsize \textrm{exc}}$, as a function of time, $\tau$. The data were taken using a condensate of 400,000 $^{87}$Rb atoms, with the Bragg detuning set to be on resonance with the measured Bragg transition at $\omega = 2\pi\times 250$ Hz for a momentum transfer given by $k=1.5\ \mu \textrm{m}^{-1}$. We expect $N_{\scriptsize\textrm{exc}}$ to go as $\tau^2$, however an interesting feature illustrated in Fig.\ \ref{fig:Bragg_v_time}  is the suppression of signal for pulses short compared to the inverse Bragg resonance, $\tau < \omega^{-1} = 0.3$ ms. For these short pulses, the associated energy uncertainty makes it impossible to resolve a $+k$ excitation from a $-k$ excitation. Photon emission from the one process cancels photon absorption from the other process. In our experiments, we did not observe Rabi flopping in the time-dependent data, which may be due to dephasing. For the measurements here, the weak beam profile used here was much smaller (7 $\mu$m $1/e^2$ waist) than that of the BEC (22 $\mu$m Thomas-Fermi radius), which complicates the Bragg response. We present this data, however, to illustrate a promising feature only available with a photon-counting approach.

\section{Conclusion}

We have demonstrated photon counting as a viable technique for Bragg spectroscopy in ultracold atoms. By measuring the response of the driving field to the atoms, one has a measurement independent of, and complementary to, the traditional atom cloud time-of-flight imaging. Our heterodyne scheme approaches shot-noise limited detection of a weak Bragg beam, which then allows us to measure the number of photons added to, or depleted from, that beam. We have shown that this measurement technique can be used simultaneously with time-of-flight imaging, and that it can probe the time dependence of the excitation process.

There are several issues to consider in applying photon-counting for Bragg spectroscopy of a BEC. Because the weak Bragg beam waist needs to be roughly matched to the transverse profile of the condensate, careful alignment must be maintained for the photon counting approach. In addition, Bragg spectroscopy for small momentum transfers can be complicated by the fact that the upward and downward portions of the lineshapes (see in Fig.\ \ref{fig:lineshape}) can begin to merge for Bragg frequencies near zero. We have also observed asymmetries between the upward and downward portions of Bragg lineshapes taken with photon counting, which we attribute to the relatively small number of $N_{\scriptsize \textrm{weak}}$ photons necessitated by the photon counting approach. We speculate that these asymmetries are related to the propagation effects inherent when the number of Bragg photons is a significant fraction of $N_{\scriptsize \textrm{weak}}$, or of the number of atoms in the condensate.

These issues may limit the usefulness of this technique to systems whose atom number is larger than the 40,000 atoms in our $^{85}$Rb BECs, where measurements of the scattered Bragg photons could open the doors to new investigations of the temperature dependent structure factor, as proposed by Stamper-Kurn \cite{Stamper-Kurn1999a}. More generally, this technique of measuring the probe in order to detect atom-light interactions could be applied to other types of ultracold atom spectroscopy as well, and it seems that the two technologies, photon counting and time-of-flight imaging, marry nicely.

We acknowledge useful conversations with J.\ Ye and the JILA ultracold atom collaboration. This work is supported by ONR and NSF.


\begin{thebibliography}{14}
\bibitem[{SA5()}]{SA5211}
\bibinfo{note}{For our photodetector, we use a Philips SA5211 Transimpedance
  amplifier}.
\end{thebibliography}

\end{document}